\documentclass[sn-mathphys,iicol,Numbered]{sn-jnl}
\usepackage{graphicx}%
\usepackage{multirow}%
\usepackage{amsmath,amssymb,amsfonts}%
\usepackage{amsthm}%
\usepackage{mathrsfs}%
\usepackage[title]{appendix}%
\usepackage{xcolor}%
\usepackage{textcomp}%
\usepackage{manyfoot}%
\usepackage{booktabs}%
\usepackage{algorithm}%
\usepackage{algorithmicx}%
\usepackage{algpseudocode}%
\usepackage{listings}%
\usepackage{type1cm}
\usepackage{anyfontsize}
\usepackage{bookmark}
\usepackage{bm}
\usepackage[T1]{fontenc}
\usepackage[utf8]{inputenc}
\usepackage{CJK}
\usepackage{subfigure}
\usepackage{multirow}

\usepackage[normalem]{ulem}

\renewcommand{\sout}{\bgroup \color{red} \ULdepth=-.5ex \ULset}

\usepackage{tikz,xcolor,hyperref}
\definecolor{lime}{HTML}{A6CE39}
\DeclareRobustCommand{\orcidicon}{
	\begin{tikzpicture}
	\draw[lime, fill=lime] (0,0) 
	circle [radius=0.16] 
	node[white] {{\fontfamily{qag}\selectfont \tiny ID}};
	\draw[white, fill=white] (-0.0625,0.095) 
	circle [radius=0.007];
	\end{tikzpicture}
	\hspace{-2mm}
}

\foreach \x in {A, ..., Z}{%
	\expandafter\xdef\csname orcid\x\endcsname{\noexpand\href{https://orcid.org/\csname orcidauthor\x\endcsname}{\noexpand\orcidicon}}
}

\begin{document}

\title[Article Title]{
Universal imprinting of short-range correlations in relativistic heavy-ion collisions
}

\author[1,2]{\fnm{Pei} \sur{Li}\orcidA{}}  \email{lip24@m.fudan.edu.cn}
\author*[1,2]{\fnm{Kai-Jia} \sur{Sun} \orcidB{}}\email{kjsun@fudan.edu.cn}
\author*[1,2]{\fnm{Bo} \sur{Zhou} \orcidC{}}\email{zhou\_bo@fudan.edu.cn}
\author*[1,2]{\fnm{Gao-liang} \sur{Ma}  \orcidD{}}\email{glma@fudan.edu.cn}

\affil[1]{\orgdiv{Key Laboratory of Nuclear Physics and Ion-beam Application (MOE), Institute of Modern Physics}, \orgname{Fudan University}, \orgaddress{\city{Shanghai}, \postcode{200433}, \country{China}}}
\affil[2]{\orgdiv{Shanghai Research Center for Theoretical Nuclear Physics}, \orgname{NSFC and Fudan University}, \orgaddress{\city{Shanghai}, \postcode{200438}, \country{China}}}

\abstract{Protons and neutrons within atomic nuclei undergo intense and fleeting encounters driven by the strong force at short distances. These interactions generate close-proximity pairs, a phenomenon known as short-range correlations (SRCs). While the properties of SRCs have been extensively studied in cold nuclear matter, their behavior under extremely hot and dense conditions remains largely unexplored. Here, we incorporate correlated nucleon configurations into relativistic heavy-ion collisions, using the quark-gluon plasma (QGP), a state of matter present microseconds after the Big Bang, as a sensitive diagnostic tool. We find that these correlations induce substantial modifications to event-by-event geometry, which are quantitatively identified through higher-order moments of the transverse profile. Most importantly, we find a surprising linear relation between QGP geometry fluctuations and the SRC scale factor spanning systems from deuteron to lead, which reflects the universal imprinting of SRCs in relativistic heavy-ion collisions. Our findings reveal the emergence of short-range structural effects across vastly different energy scales from low-energy electron scattering to high-energy nuclear collisions.
}


\maketitle
\section{Introduction}

\begin{figure*}[!t]
\centering 

\includegraphics[scale=0.2146]{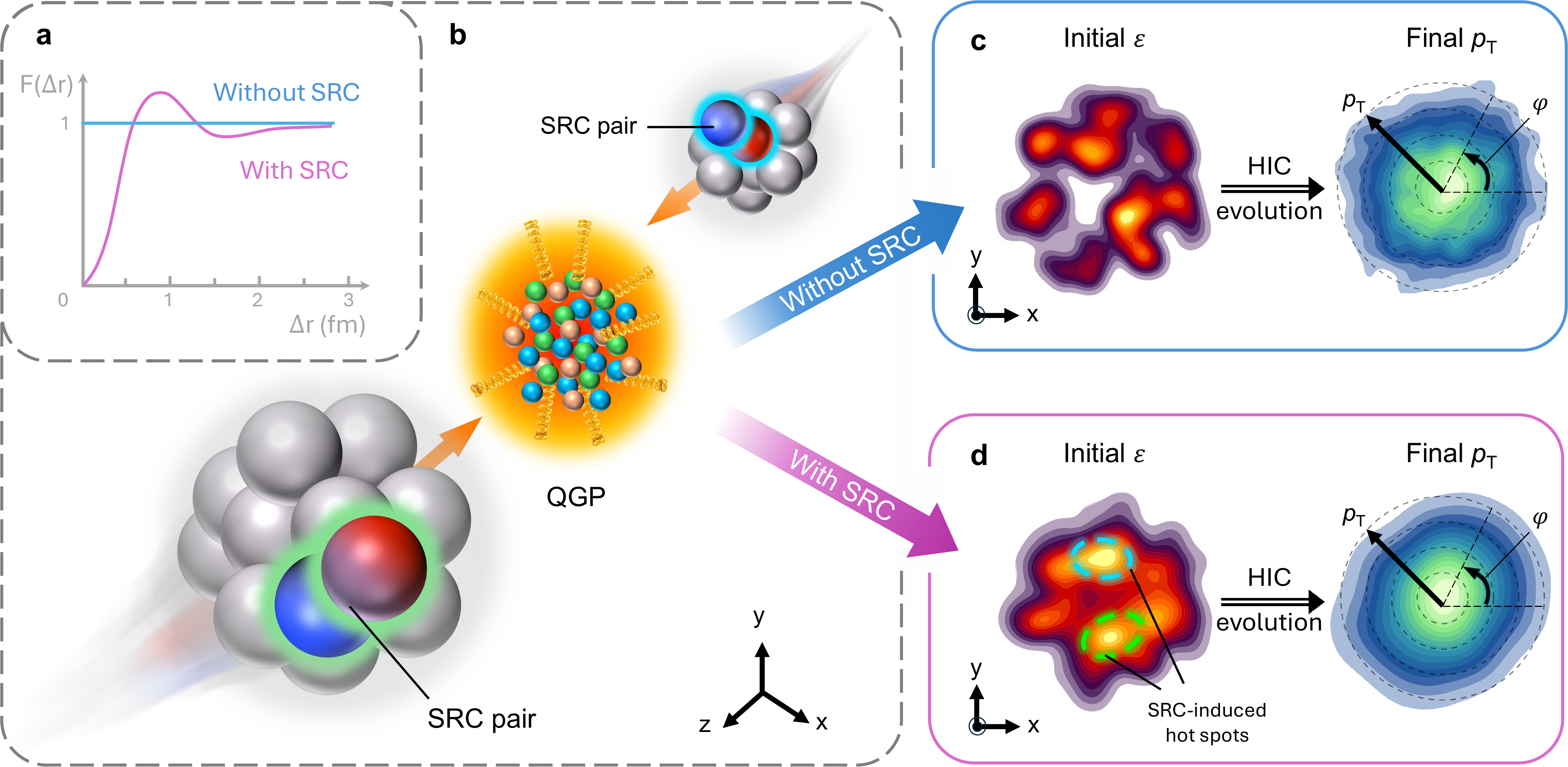}
\caption{
\textbf{Schematic representation for heavy-ion collisions with NN SRCs.} \textbf{a}, Example of nucleon-nucleon correlation function ($F(\Delta r)$) with (pink line) or without (blue line) SRC as a function of their relative distance. \textbf{b}, The collision of two nuclei generates an expanding fireball of quark-gluon plasma. \textbf{c,d}, The underlying nucleon-nucleon interactions modulate the fluctuations in the initial-state transverse energy density ($\varepsilon$) profile. Nucleon pairs at short distances act as localized hot spots encoded into the initial geometry. Through hydrodynamic evolution, these initial-state fluctuations are mapped onto the final-state transverse momentum ($p_T$) distributions of produced hadrons.
}
\label{Fig:diagram}
\end{figure*}

Atomic nuclei are often described as collections of protons and neutrons moving independently within an average potential. However, this simple picture is incomplete. At short distances, nucleons experience strong interactions that momentarily drive them into close proximity, generating pairs with exceptionally high momentum~\cite{Arrington:2011xs,Hen:2012fm,Ye:2013wih,Hen:2016kwk,CLAS:2019vsb,Li:2022fhh,Shang:2025vtd}. These fleeting configurations, known as short-range correlations (SRCs), arise from the most intense components of the nuclear force and represent a fundamental aspect of the nuclear structure. Over the past decades, electron scattering experiments have shown that such correlated pairs represent a universal feature spanning from light nuclei to heavy systems~\cite{Subedi:2008zz,Hen:2014yfa,CLAS:2018yvt,Li:2018lpy,Cruz-Torres:2019fum,Cai:2025txx,Tian:2026wkr,Nguyen:2026nat}. Despite these progresses, SRCs remain difficult to interpret intuitively: they are transient, localized, and embedded within the many-body system. 

Experimental studies of SRCs have relied on quasi-elastic electron scattering (QES) measurements, which have revealed characteristic scaling plateaus in cross-section ratios, showing that nucleons with momenta above the Fermi surface predominantly originate from correlated pairs~\cite{Frankfurt:1988nt,Fomin:2011ng}. More exclusive measurements, in which both nucleons from a correlated pair are detected, further established the strong dominance of neutron--proton pairs~\cite{Subedi:2008zz,LabHallA:2014wqo,CLAS:2018yvt,Yankovich:2024xze}. However, electron-scattering experiments probe nuclei under a kinematically restricted regime, where the extraction of SRC signals often relies on specific kinematic selections and substantial corrections associated with final-state interactions and reaction mechanisms.

To circumvent limitations inherent in low-energy electron scattering, we utilize relativistic heavy-ion collisions (HIC) to investigate nuclear SRCs. In these collisions, two nuclei collide at nearly the speed of light and create quark-gluon plasma (QGP) matter with temperatures exceeding trillions of degrees ($\sim 10^{12}\ \mathrm{K}$), whose evolution is highly sensitive to the initial spatial geometry of the nuclei~\cite{ALICE:2018lao,Zhang:2021kxj,ALICE:2021gxt,ATLAS:2022dov,STAR:2024wgy,Zhao:2024lpc}. Previous studies have concentrated on global properties such as nuclear sizes and shapes~\cite{Ollitrault:1992bk,Alver:2010gr,Bilandzic:2010jr,PhysRevC.86.044908,Bilandzic:2013kga,PhysRevC.93.024907,Jia:2021tzt,Jia:2021qyu,Jia:2022ozr}. In contrast, in relativistic heavy-ion collisions (Fig.~\ref{Fig:diagram}b), SRC-induced localised high-density fluctuations, originating from the interplay of the repulsive core and the attractive tensor force between nucleons (Fig.~\ref{Fig:diagram}a), modify the event-by-event initial-state energy-density $\varepsilon$ profile (left panel of Fig.~\ref{Fig:diagram}d), compared to that without SRC (left panel of Fig.~\ref{Fig:diagram}c). These modifications are then carried through the hydrodynamic expansion and leave observable imprints on the final-state transverse momentum $p_{\rm T}$ distributions (right panels of Fig.~\ref{Fig:diagram}c and d), thus providing a complementary probe of SRC in the extreme high-energy environment.

In this work, we investigate the impact of SRCs on the local properties of QGP in relativistic heavy-ion collisions. We develop an analytical two-point correlation function to study the influence of attractive and repulsive NN potentials on initial-state fluctuations and find that higher-order cumulants are more sensitive. Moving beyond the analytical approach, we incorporate these SRC-induced modifications into a full-fledged realistic hydrodynamic model with many-body sampling and find clear imprints on final-state observables, most prominently in the size-related cumulants. By mapping the relative ratio to the SRC scale factor from electron scattering, we establish a universal linear relation across nuclear species. This work points towards a unified description of quantum chromodynamic matter across vastly different energy scales, showing that the same underlying interactions govern both static nuclear structures and rapidly evolving quark-gluon systems.

\begin{figure*}[ht]
\centering 
\includegraphics[scale=0.83]{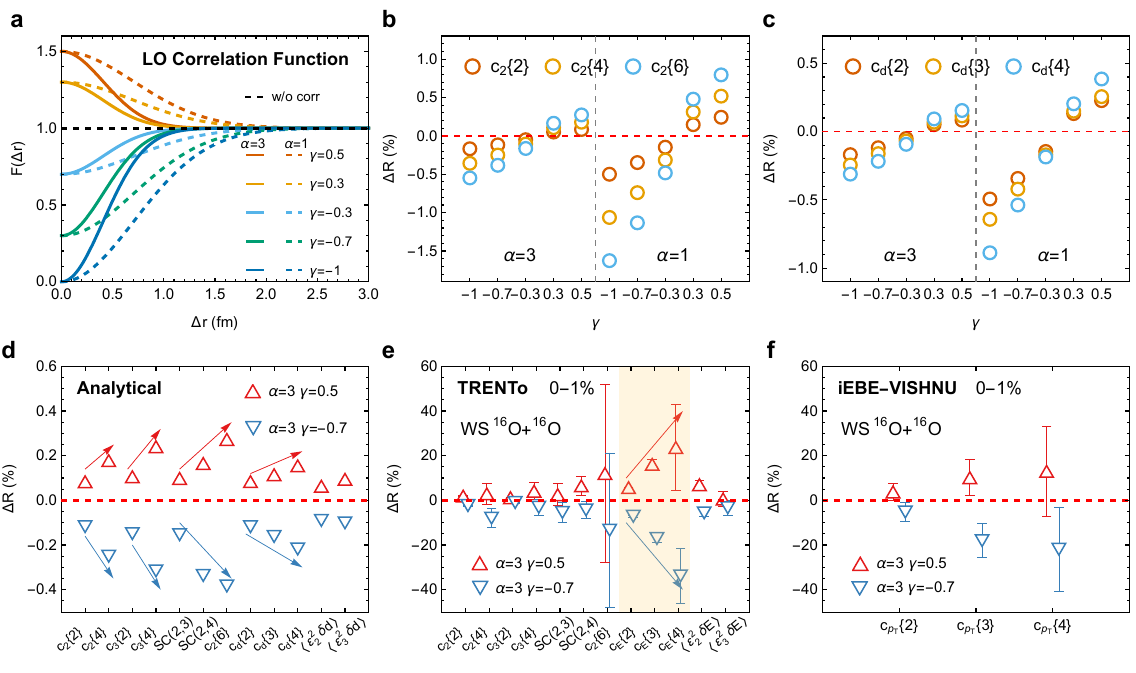} 
\caption{
\textbf{SRC-induced fluctuations in relativistic nuclear collisions.} \textbf{a}, The nucleon distributions from NN correlations under the parameterized leading-order two-point correlation function. \textbf{b,c}, Analytical results for correlation effects on eccentricity cumulants (\textbf{b}) and inverse transverse size cumulants (\textbf{c}) in ultracentral $^{16}$O+$^{16}$O collisions. \textbf{d}, Analytical results for correlation effects in various observables in ultracentral $^{16}$O+$^{16}$O collisions with a tetrahedral 4$\alpha$-cluster structure under attractive (red) and repulsive (blue) leading-order correlation functions. \textbf{e}, Same as in \textbf{d}, but for initial-state T\raisebox{-0.5ex}{R}ENTo model results using a Woods-Saxon distribution. \textbf{f}, The relative ratio $\Delta R$ for mean transverse momentum cumulants $c_{p_{\mathrm{T}}}\{n\}$ with NN correlations using final-state iEBE-VISHNU hydrodynamic model.
}
\label{Fig:correlationF} 
\end{figure*}

\section{Results}
\bmhead{Decoding SRCs via fluctuations in HIC}The nucleon-nucleon(NN) correlations originate from both statistical effects (e.g., Pauli repulsion) and dynamical effects due to the nature of the nuclear force. Incorporating all necessary many-body correlation operators into a precise Hamiltonian leads to computationally intensive calculations. To simplify this, two-point correlation functions $F(\boldsymbol{r}_1,\boldsymbol{r}_2)$ are commonly used to describe NN correlations~\cite{Cruz-Torres:2017sjy,Alvioli:2009ab},
\begin{equation}
\begin{aligned}
F(\boldsymbol{r}_1,\boldsymbol{r}_2)=\frac{A}{A-1}\frac{\rho^{(2)}(\boldsymbol{r}_1,\boldsymbol{r}_2)}{\rho^{(1)}(\boldsymbol{r}_1)\rho^{(1)}(\boldsymbol{r}_2)}.
\label{Eq:1}
\end{aligned}
\end{equation}
Evidently, the two-point correlation function depends on the interaction adopted in many-body calculations~\cite{CLAS:2020rue,CLAS:2020mom}, which can be parametrized as~\cite{Cruz-Torres:2017sjy},
\begin{equation}
\begin{aligned}
F(\Delta r)=1+e^{-\alpha(\Delta r)^2}(\gamma+\sum_{i}^{3}\beta_i(
\Delta r)^{i+1}),
\label{Eq:2}
\end{aligned}
\end{equation}
where $\Delta r=|\boldsymbol{r}_1-\boldsymbol{r}_2|$ denotes the relative distance between the two nucleons. Parameters $\alpha$ and $\gamma$ are used to control the nature of the interaction. Specifically, $\gamma>0$ or $\gamma<0$ indicate an attractive or repulsive central potential, respectively, and $\gamma\rightarrow0$ or $\alpha\rightarrow\infty$ means that there are no correlations. Higher-order expansion terms with $\beta_i$ correspond to medium- and long-range properties.

To elucidate the general characteristics of initial-state fluctuations induced by short-range correlations, we deploy an analytical theoretical framework using smooth density~\cite{Li:2025hae}, in which only non-spherical nuclear structures exhibit anisotropy due to the neglect of nucleon position fluctuations. The eccentricities $\varepsilon_{n}$ \cite{Bilandzic:2010jr,PhysRevC.83.064904,Moreland:2014oya,Jia:2021tzt} and the inverse transverse size $d_{\perp}$ \cite{Jia:2021tzt,Jia:2021qyu} are typically used to describe the initial geometry and size of the QGP, respectively. We choose a regular triangular pyramid $4\alpha$-cluster structure of oxygen nuclei to characterize the one-body density~\cite{Li:2020vrg,Horiuchi:2023pym}, which theoretically possesses strong anisotropy to manifest changes more significantly resulting from the correlation effect alone. See Methods for details.

The leading-order correlation functions $F(\Delta r)=1+\gamma e^{-\alpha(\Delta r)^2}$\cite{Cruz-Torres:2017sjy} are used to facilitate analytical integral calculations, as shown in Fig.~\ref{Fig:correlationF}a. We quantify the relative sensitivity induced by NN SRCs in heavy-ion collisions via the relative ratio
\begin{equation}
\begin{aligned}
\Delta R(\mathcal{Q}) = \mathcal{Q}_{\mathrm{cor}}/\mathcal{Q}_{\mathrm{uncor}} - 1,
\label{Eq:3}
\end{aligned}
\end{equation}
where $\mathcal{Q}$ is an initia-state cumulant. Figure.~\ref{Fig:correlationF}b and c depicts the behavior of $\Delta R$ for both eccentricity cumulants $c_2{\{n\}}$ and inverse transverse size cumulants $c_d{\{n\}}$ of different orders under the modulation of correlation parameters $\alpha$ and $\gamma$. When $\gamma>0$, the attractive interaction induces an enhancement in the initial-state cumulants of various orders, due to the increased probability of hot spots and larger fluctuations caused by SRCs, while the repulsive interaction triggers suppression when $\gamma<0$. Simultaneously, a smaller $\alpha$ signifies a broader spatial range of the interaction, and the resulting enhancement or suppression of the initial-state fluctuations becomes considerably more pronounced. Higher-order cumulants display strikingly enhanced sensitivity to NN correlations. Thus we can identify two quantitative hierarchies:
\begin{equation}
\begin{aligned}
|\Delta R(c_2\{6\})|>|\Delta R(c_2\{4\})|>|\Delta R(c_2\{2\})|,
\label{Eq:4}
\end{aligned}
\end{equation}

\begin{equation}
|\Delta R(c_d\{4\})|>|\Delta R(c_d\{3\})|>|\Delta R(c_d\{2\})|.\ 
\label{Eq:5}
\end{equation}

All commonly used cumulants for the two cases with $\alpha=3$ in the analytical framework are shown in Fig.~\ref{Fig:correlationF}d.
To incorporate a more realistic description of the QGP, which includes accounting for realistic nucleon-nucleon collision cross-sections, introducing spatial fluctuations of nucleon positions, and embedding sub-nucleon structures, we next deploy a hydrodynamic Monte Carlo framework to simulate the whole evolution of heavy-ion collisions at energy $\sqrt{s_{NN}} = 5.36$~TeV. The model parameters are consistent with those in Ref.~\cite{Hujy:2025eid}. To introduce nucleon correlations more reasonably, we employ many-body sampling via the Schrödinger Generator~\cite{Jiang2026}, which serves as the initial nuclear configuration input. See
Methods for details. The system size is characterized by the ratio of total energy and total entropy $E/S$ (marked with $E$) rather than inverse transverse size $d_{\perp}$ (marked with $d$) in the initial-state T\raisebox{-0.5ex}{R}ENTo model, due to its more direct relevance to the initial energy density~\cite{Giacalone:2020dln}.

\begin{figure*}[ht]
\centering
\includegraphics[scale=0.83]{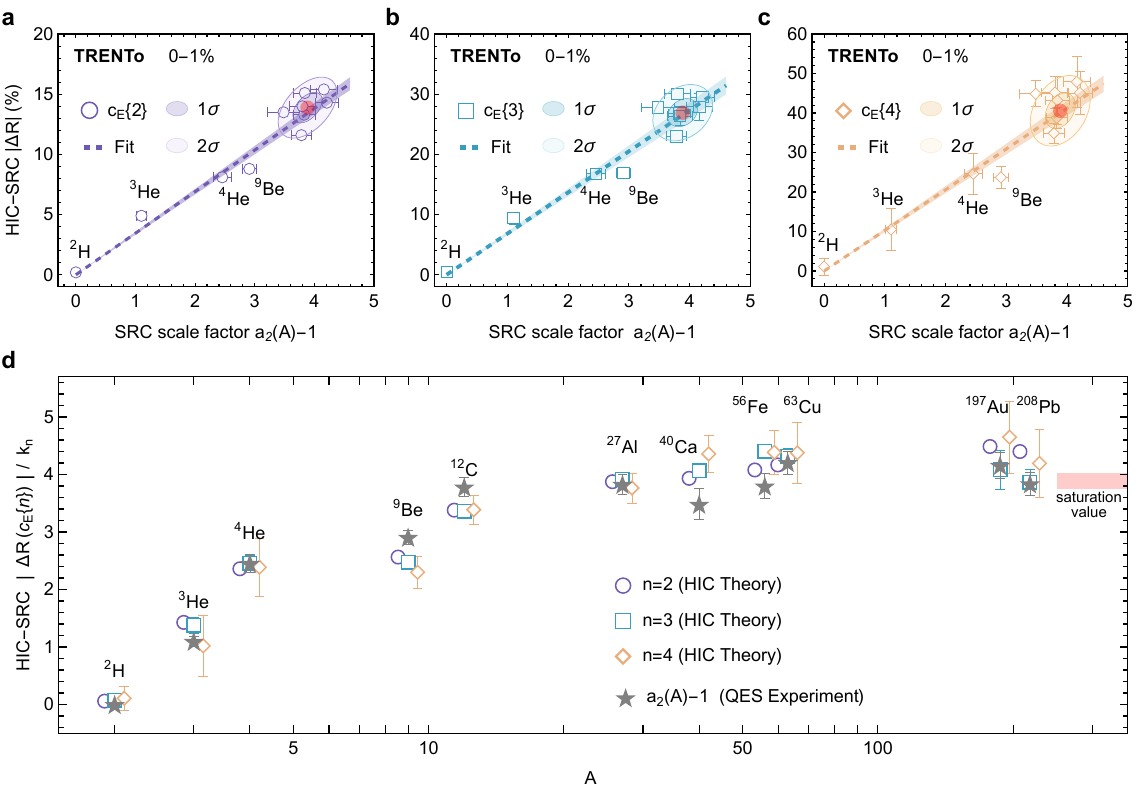}
\caption{\textbf{Linear relation of SRCs between heavy-ion collisions and quasi-elastic electron scatterings.} \textbf{a,b,c}, The linear relation between relative ratio $|\Delta R(c_{E}\{n\})|$ (n=2, 3, 4) in $A+A$ collisions on the SRC scale factor $a_2(A)-1$ in quasi elastic scattering experiments for different nuclei, i.e., $|\Delta R(c_{E}\{n\})|=k_{n}\left[a_2(A)-1\right]$. The correlation function of AV8$^{\prime}$ interaction~\cite{Alvioli:2009ab} is used for all nuclei in T\raisebox{-0.5ex}{R}ENTo model. The results of SRC scale factor represent the updated experimental electron scattering data from Jefferson Lab~\cite{CLAS:2019vsb,Zhang:2025nst}. The error bars on the symbols show statistical uncertainties, the uncertainties associated with the linear fitting denote the 1$\sigma$ uncertainty or 68$\%$ confidence level. For medium-to-heavy nuclei with $A \ge 12$, a standard deviation ellipse fitting is utilized in the saturation region, and the inner and outer elliptical domains represent the 1$\sigma$ and 2$\sigma$ uncertainties yielded. \textbf{d}, The scaled relative ratio (open symbols) and SRC scale factor (solid star) are shown as functions of nucleon number $A$ from $^{2}\mathrm{H}$ to $^{208}\mathrm{Pb}$. The red shaded region indicates an averaged effective SRC scale factor of $4.89 \pm 0.14$ by integrating the three distinct orders of size-related fluctuation cumulants.
}
\label{Fig:scaling} 
\end{figure*}

Figure~\ref{Fig:correlationF}e presents the initial-state T\raisebox{-0.5ex}{R}ENTo model results for ultracentral 0--1$\%$ range. The correlation effect revealed by the T\raisebox{-0.5ex}{R}ENTo simulations agrees with the analytical trend, which faithfully reproduces the quantitative hierarchies that are formally established in Eqs.~(\ref{Eq:2}) and~(\ref{Eq:3}). The inclusion of nucleon-position fluctuations directly amplifies fluctuations in the initial-state energy density, which renders the correlation effect in Monte Carlo model far more prominent than in analytical calculations by elevating the expected signal from less than 1$\%$ to anywhere between several percent and tens of percent. The correlation effect on shape $\varepsilon_n$ is very limited, as reflected in different three-point cumulants:
\begin{equation}
|\Delta R(c_{E}\{3\})|>|\Delta R(\langle \varepsilon_2^2 \delta E\rangle)|\approx |\Delta R(\langle \varepsilon_3^2 \delta E\rangle)|.
\label{Eq:6}
\end{equation}
The pure size cumulants exhibit a substantially stronger sensitivity to correlation effects than the mixed shape-size cumulants. For the 3- and 4-point correlators $c_{E}\{3\}$ and $c_{E}\{4\}$, the SRC effect remains statistically significant with relative deviations $\Delta R$ exceeding $10\%$. The linear response in hydrodynamic evolution, $p_{\mathrm{T}} \propto E/S$, enables us to probe initial-state size fluctuations using final-state observables~\cite{Giacalone:2020lbm,Nielsen:2023znu,Zhang:2024vkh}. Fig.~\ref{Fig:correlationF}f presents the results of the iEBE-VISHNU hydrodynamic model for the higher-order cumulants of the final-state mean transverse momentum in ultracentral collisions. The pronounced impact of SRC effect on size-dependent initial cumulants successfully maps onto the corresponding final-state transverse momentum cumulants $c_{p_{\mathrm{T}}}\{n\} (n\geq 2)$ , which then emerge as promising candidates to probe the SRC effect in heavy-ion collisions. This also indicates that nucleon-nucleon correlations are essential for a complete description of higher-order final-state transverse momentum fluctuations~\cite{ALICE:2025rtg}.

\bmhead{Unveiling universality of HIC-SRC effect} 
In quasi-elastic electron scattering experiments at Jefferson Lab, the ratio of the inclusive scattering cross-section per-nucleon for nucleus $A$ to deuterium, at large four-momentum transfer of $Q^{2} > 1.4\ \text{GeV}^{2}$ and within the Bjorken $x_B$ regime of $1.5 \le x_B < 2$, defines the SRC scale factor, $a_{2}(A) = \frac{2\sigma_{A}}{A\sigma_{d}}$. This dimensionless quantity serves as a direct measure that quantifies the intrinsic strength of the SRC effects.

To establish a direct link between the SRC scale factor and the features of fluctuations in heavy-ion collisions, we systematically investigate the HIC-SRC effect in $A+A$ collisions across all nuclear species for which experimental data on SRC scale factors are currently available. We use the realistic correlation function by fitting parameters in Eq.~(\ref{Eq:2}) with the AV8$'$ interaction from Ref.~\cite{Alvioli:2009ab}. Regarding the one-body density distributions, standard Woods Saxon profiles are adopted for $^4\text{He}$ and heavier systems ($A \geq 12$), and the harmonic-oscillator distributions are implemented for $^3\text{He}$ and $^9\text{Be}$~\cite{DeJager:1974liz,DeVries:1987atn}, while deuterium $^{2}\text{H}$ serves as the baseline of NN correlation that is explicitly modeled by two localized nucleon wave functions.

Figure~\ref{Fig:scaling}a--c reveals the relation between the relative ratio $\Delta R(c_{E}\{n\})$ in the ultracentral range and the SRC scale factor from $^{2}\mathrm{H}$ to $^{208}\mathrm{Pb}$. A universal linear relation is consistently observed across the initial-state size fluctuations of various orders, $|\Delta R(c_{E}\{n\})|=k_{n}\left[a_2(A)-1\right]$, where the proportionality coefficient $k_{n}$ acts as a direct measure of the relative sensitivity of heavy-ion observables to the SRC dynamics, which essentially quantifies the amplification efficiency of these sub-femtometer structures as they evolve through the relativistic collision environment. In general, these proportionality coefficients $k_{n}$ are dependent on multiple factors, including collision energy, order of fluctuations, selected centrality window, and nature of the NN interaction. To isolate the targeted correlation effects, we operate under the well-justified assumption that all such collision-related conditions remain identical across different $A+A$ systems. In this way, the observed variations stem solely from the variations in the initial nuclear configurations. Employing a standard deviation ellipse fitting for medium-to-heavy nuclei with nucleon number $A \ge 12$, one can observe that the scaled relative ratio and the SRC scale factor maintain a robust positive correlation even within the saturation region, which confirms that the underlying physical mapping remains intact as nuclear systems reach the near-saturation density regime.

Figure~\ref{Fig:scaling}d shows the scaled relative ratios $\Delta R(c_{E}\{n\})/k_n$ and the SRC scale factor $a_2(A) -1$ as functions of the nucleon number $A$. The scaled relative ratios across various orders are found to be consistent with results from low-energy electron-scattering experiments, while the adoption of spherical nucleon distributions represents an oversimplified assumption for certain species. For the nuclei $^{9}\text{Be}$ and $^{12}\text{C}$, $\textit{ab initio}$ calculations substantiate the presence of a pronounced cluster structure in the ground state\cite{Arai:1996dq,Huang:2021mct,Horiuchi:2023pym}, which enhances the local nucleon density to favor short-range correlations.

\section{Discussion}

By incorporating short-range correlations via a two-point correlation function, we show that higher-order initial-state cumulants of nuclear size sensitively capture attractive and repulsive nucleon-nucleon interactions. Hydrodynamic modeling confirms these initial geometric modifications persist through expansion, surfacing as clear imprints on final-state size-related cumulants. Crucially, mapping the relative ratio $\Delta R$ against the electron-scattering scale factor $a_2(A)$ reveals a universal linear scaling across systems from light $^2\text{H}$ to heavy $^{208}\text{Pb}$. This scaling across vastly different energy scales, from cold ground states in electron scattering to rapidly expanding hot matter in heavy-ion collisions, confirms that the measurable imprints of SRCs are resistant to the extreme conditions of high-energy collisions. These results establish SRCs as an integral part of initial collision geometry that shapes the emergent phenomena of the QGP, bridging microscopic nuclear structure with macroscopic collective dynamics.

Our work opens new avenues for future experimental campaigns aimed at scanning diverse light nuclear species at the Large Hadron Collider (LHC) and the Super Proton Synchrotron (SPS)~\cite{Triantafyllou:2025jve}. Beyond the collider experiments, the linear scaling established here also has potential implications for constraining the nuclear equation of state in dense astrophysical environments such as neutron stars, bridging the gap between laboratory nuclear physics and astrophysical observations~\cite{Danielewicz:2002pu,Sedrakian:2024uma,Fomin:2026swt}. 

\section{Methods} 
\bmhead{Analytical framework for NN-correlated calculations} 
The nuclear structure, including its shape, clustering and size, can be probed via high-energy heavy-ion collisions, based on analytical calculations of the one-body density distribution~\cite{Jia:2021qyu,Jia:2021tzt,Li:2025hae}. A typical one-body nucleon density distribution is obtained by integrating the $A$-body nucleon density over $A-1$ dimensions:
\begin{equation}
\begin{aligned}
\rho^{(1)}(\boldsymbol{r}_1) = A\int |\Phi(\boldsymbol{r}_1,\boldsymbol{r}_2,...,\boldsymbol{r}_A)|^2 \mathrm{d}\boldsymbol{r}_2\mathrm{d}\boldsymbol{r}_3...\mathrm{d}\boldsymbol{r}_A,
\label{Eq:7}
\end{aligned}
\end{equation}
where $\Phi(\boldsymbol{r}_1,\boldsymbol{r}_2,...,\boldsymbol{r}_A)$ is $A$-body wave function and the square of the wave function $|\Phi|^2$ represents the $A$-body density distribution. However, the one-body density obviously does not contain the natural NN correlations. The existing correlations in previous studies on nuclear structure in heavy-ion collisions were simply brought about by the Wigner rotation integral~\cite{Giacalone:2023hwk,Mehrabpour:2025ogw}. To include the lowest-order NN correlations, the two-body nucleon density can be obtained by integrating over $A-2$ dimensions~\cite{Kanada-Enyo:2011jkr}:
\begin{equation}
\begin{aligned}
\rho^{(2)}(\boldsymbol{r}_1,\boldsymbol{r}_2)&= A(A-1)\\
&\times\int |\Phi(\boldsymbol{r}_1,\boldsymbol{r}_2,...,\boldsymbol{r}_A)|^2\mathrm{d}\boldsymbol{r}_3\mathrm{d}\boldsymbol{r}_4...\mathrm{d}\boldsymbol{r}_A.
\label{Eq:8}
\end{aligned}
\end{equation}
If we consider the lowest-order two-body NN correlation, then it satisfies $\rho^{(2)}(\boldsymbol{r}_1,\boldsymbol{r}_2) \neq [(A-1)/A]\rho^{(1)}(\boldsymbol{r}_1)\rho^{(1)}(\boldsymbol{r}_2)$. The two-body density distribution and the one-body density distribution are related. Thus, the effective one-body density $\rho^{(1)}_{\text{cor}}$, which incorporates NN correlations, is given by
\begin{equation}
\begin{aligned}
\rho^{(1)}_{cor}(\boldsymbol{r}_1)=\frac{1}{A-1}\int \rho^{(2)}(\boldsymbol{r}_1,\boldsymbol{r}_2)\mathrm{d}\boldsymbol{r}_{2},
\label{Eq:9}
\end{aligned}
\end{equation}
where $\rho^{(2)}(\boldsymbol{r}_1,\boldsymbol{r}_2)=\left[\left(A-1\right)/A\right]F(\boldsymbol{r}_1,\boldsymbol{r}_2)\rho^{(1)}(\boldsymbol{r}_1)\rho^{(1)}(\boldsymbol{r}_2)$ using Eq.~(\ref{Eq:1}). Building on the framework for $^{20}$Ne in Ref.~\cite{Li:2025hae}, we extend the approach to evaluate expectation values via the effective one-body density, which equivalently performs the two-body integration incorporating the NN correlation function.

Following Ref.~\cite{Bozek:2019wyr,YuanyuanWang:2024sgp}, we represent $^{16}$O in analytical calculations by a frozen cluster structure with the approximate one-body density $\rho^{(1)}(\boldsymbol{r})=4\sum_{i}^{4}\rho_{\alpha_i}(\boldsymbol{r},\boldsymbol{R}_i)$, where $\rho_{\alpha_i}(\boldsymbol{r},\boldsymbol{R}_i)=(\frac{1}{\pi b^2})^{\frac{3}{2}}\ \text{exp}\left[{-\frac{1}{b^2}(\bm{r}-\bm{R}_i)^2} \right]$ is the distribution of the
nucleons in the $i$-th alpha cluster, $\bm{R}_i$ is the center-of-mass position of the $i$-th cluster, and $b$ is the harmonic oscillator size. By setting the harmonic oscillator size parameter $b=1.76$ fm, the resulting root-mean-square (rms) radius of $^{16}$O is consistent with the experimental charge radius (2.7 fm).

Using the effective one-body density $\rho^{(1)}$, the total initial nucleon density in the 2D transverse plane in ultracentral HICs is defined as the arithmetic average of the projectile and target densities,
\begin{equation}
\begin{aligned}
\rho(\bm{r_{\perp}})=\int[\mathcal{R}_{p}\rho_{cor}^{(1)}(\bm{r})+\mathcal{R}_{t}\rho_{cor}^{(1)}(\bm{r})]/2\ \mathrm{d}z,
\label{Eq:A0}
\end{aligned}
\end{equation}
where $\bm{r}=(r_{\perp},\phi,z)$ and $\bm{r_{\perp}}=(r_{\perp},\phi)$. The $\mathcal{R}_{p}$ and $\mathcal{R}_{t}$ are the Euler rotation matrices for the projectile and the target nuclei, respectively. The initial-state quantity $\mathcal{Q}$ can be expressed as
\begin{equation}
\begin{aligned}
\mathcal{Q}&=\langle q(\bm{r_{\perp}})\rangle=\int q(\bm{r_{\perp}})\rho(\bm{r_{\perp}}) \mathrm{d}\bm{r_{\perp}}\mathrm{d}\Omega_{p}\mathrm{d}\Omega_{t}\\
&=\int \mathcal{R}_{p,t}(\mathcal{I}_{p}+\mathcal{I}_{t})/2\ \mathrm{d}\Omega_{p,t},
\label{Eq:A1}
\end{aligned}
\end{equation}
where 

\begin{equation}
\begin{aligned}
\mathcal{I}_{p,t}&=\int q(\bm{r_{\perp}})\rho^{(1)}_{cor}(\bm{r})\mathrm{d}\bm{r}=\mathcal{I}^{(1)}_{p,t}-\mathcal{I}^{(2)}_{p,t}.
\label{Eq:A2}
\end{aligned}
\end{equation}
Here, we use the relationship $F(\bm{r},\bm{r}^{\prime})=1-C(\bm{r},\bm{r}^{\prime})$, and the expected value can be divided into two parts. The one directly generated by the original one-body distribution $\mathcal{I}^{(1)}_{p,t}$ and the other produced by the two-body correlation distribution $\mathcal{I}^{(2)}_{p,t}$:

\begin{align}
  \mathcal{I}^{(1)}_{p,t} &= \int q(\bm{r_{\perp}})\rho^{(1)}(\bm{r})\,\mathrm{d}\bm{r},
  \label{Eq:A3} \\
  \begin{split}
    \mathcal{I}^{(2)}_{p,t} &= \frac{1}{A}\int_{\bm{r}\neq \bm{r}^{\prime}} q(\bm{r_{\perp}}) C(\bm{r},\bm{r}^{\prime}) \\
    &\quad \times \rho^{(1)}(\bm{r})\rho^{(1)}(\bm{r}^{\prime})\,\mathrm{d}\bm{r}\,\mathrm{d}\bm{r}^{\prime}.
  \end{split}
  \label{Eq:A4}
\end{align}

\bmhead{Cumulants in HICs}
\label{quantities}
The eccentricity $\varepsilon_{n}$ in the center-of-mass frame is typically used to describe the initial geometrical condition of collisions via 2D multi-pole expansion. The eccentricity vectors are defined as~\cite{PhysRevC.83.064904,Moreland:2014oya},
\begin{align}
\varepsilon_{n}e^{in\Phi}=-\frac{\langle r_{\perp}^{n}e^{in\phi}\rangle}{\langle r_{\perp}^{n}\rangle},\ \ \ \ \ n\geq 2.
\label{Eq:A5}
\end{align}
Additionally, the size of the initial state can be quantified by the inverse transverse size~\cite{Li:2025hae},
\begin{equation}
d_{\perp}=(\langle x^2y^2\rangle-\langle xy\rangle^2)^{-1/4}.
\label{Eq:A6}
\end{equation}
With the definitions of nuclear shape and size, we can perform a more refined quantification of high-order fluctuations, e.g. the expressions for eccentricity cumulants~\cite{Bhalerao:2006tp,Voloshin:2007pc} of any order:
\begin{align}
  c_{n}\{2\}&=\langle \varepsilon_{n}^{2}\rangle, \label{Eq:A7} \\
  c_{n}\{4\}&=\langle \varepsilon_{n}^{4}\rangle-2\langle\varepsilon_{n}^{2}\rangle^{2}, \label{Eq:A8} \\
  c_{n}\{6\}&=(\langle \varepsilon_{n}^{6}\rangle-9\langle\varepsilon_{n}^{4}\rangle\langle\varepsilon_{n}^{2}\rangle+12\langle\varepsilon_{n}^{2}\rangle^3)/4, \label{Eq:A9}
\end{align}

and the expressions for the inverse transverse size cumulants~\cite{Jia:2021qyu,Nielsen:2023znu} of any order:

\begin{align}
  c_{d}\{2\}&=\langle(\frac{\delta d_{\perp}}{ d_{\perp}})^2\rangle, \label{Eq:A10} \\
  c_{d}\{3\}&=\langle(\frac{\delta d_{\perp}}{ d_{\perp}})^3\rangle, \label{Eq:A11} \\
  c_{d}\{4\}&=\langle(\frac{\delta d_{\perp}}{ d_{\perp}})^4\rangle-3\langle(\frac{\delta d_{\perp}}{ d_{\perp}})^2\rangle^{2}. \label{Eq:A12}
\end{align}

In addition, we also consider mixed cumulants, including symmetric cumulants~\cite{Bilandzic:2010jr,Bilandzic:2013kga},
\begin{equation}
SC(m,n)=\langle \varepsilon_{m}^{2}\varepsilon_{n}^{2}\rangle-\langle\varepsilon_{m}^{2}\rangle\langle\varepsilon_{n}^{2}\rangle,
\label{Eq:A13}
\end{equation}
and the covariance correlation coefficient~\cite{Bozek:2016yoj}:
\begin{equation}
\mathrm{cov}(\varepsilon_{n}^{2}, \delta d_{\perp})=\langle \varepsilon_{n}^{2}\delta d_{\perp}\rangle.
\label{Eq:A14}
\end{equation}
Following the definition of inverse transverse size cumulants, we construct equivalent observables using $\langle p_{\mathrm{T}}\rangle$ in the final state and the ratio of total energy and total entropy $E/S$ in the initial state. The established scaling $p_{\mathrm{T}} \propto E/S \propto d_{\perp}$ ensures that these cumulants effectively capture the same geometric information at different stages of the collision~\cite{Giacalone:2020dln,Nielsen:2023znu}. 

\bmhead{Many-body sampling with the Schr\"{o}dinger Generator} We expect to investigate the NN correlation effect while maintaining the identical one-body density distribution, thus avoiding interference from mean field effects. Sampling is performed using the following formula:
\begin{equation}
\rho^{(A)}_{cor}(\boldsymbol{r}_1,\boldsymbol{r}_2,...,\boldsymbol{r}_A) =\frac{1}{\mathcal{N}}|\Phi(\boldsymbol{r}_1,\boldsymbol{r}_2,...,\boldsymbol{r}_A)|^2\prod_{i< j}F (\boldsymbol{r}_i,\boldsymbol{r}_j),
\label{Eq:10}
\end{equation}
where $\Phi(\boldsymbol{r}_1,\boldsymbol{r}_2,...,\boldsymbol{r}_A)$ is antisymmetric under exchange of identical fermions and includes the statistical correlations caused by the Pauli exclusion principle, and $\prod_{i<j}F (\boldsymbol{r}_i,\boldsymbol{r}_j)$ includes the dynamical correlations arising from the nature of the NN interactions. However, directly implementing the full correlated distribution in Eq.~(\ref{Eq:10}) poses significant computational challenges for heavy nuclei. To ensure computational tractability while capturing the essential short-range dynamics, we focus on the leading-order two-body corrections for nuclei with a Woods Saxon (WS) distribution $\rho(r)=\rho_0(1+wr^2/R_0^2)/(1+\exp[(r-R_0)/a])$, the following formula is employed for sampling:
\begin{equation}
\rho^{(A)}_{WS,cor}(\boldsymbol{r}_1,\boldsymbol{r}_2,...,\boldsymbol{r}_A) =\frac{1}{\mathcal{N}}\prod_{i}^{A}\rho^{(1)}_{WS}(\boldsymbol{r}_i)\prod_{i< j}F (\boldsymbol{r}_i,\boldsymbol{r}_j).
\label{Eq:11}
\end{equation}
which preserves the desired one-body density while incorporating two-body correlations.

Accurate sampling of such high-dimensional correlated distributions remains a longstanding challenge in high-energy nuclear physics~\cite{Kersevan:2004yg,Alvioli:2009ab,Huang:2025uvc}. Conventional Markov Chain Monte Carlo (MCMC) methods~\cite{Hastings:1970aa,Alvioli:2009ab} often suffer from poor efficiency in high dimensions.

To overcome this limitation, we employ a recently developed Schr\"{o}dinger Generator for high-dimensional sampling and integration~\cite{Jiang2026}. This approach decomposes the transformation Jacobian into two complementary components: (i) an adaptive mapping that reduces estimator variance by learning marginal structures in each dimension, and (ii) a normalizing-flow-based transformation that captures non-factorizable correlations in the target distribution. A final resampling step ensures unbiased sampling even when the learned transformation is not exact. 

This method exhibits stable and scalable performance for strongly correlated quantum many-body systems in dimensions up to several hundred~\cite{Jiang2026}. We have verified that the generated samples faithfully reproduce the imposed NN correlations.

\backmatter

\bmhead{Acknowledgments}
Discussions with Zhi-hong Ye, Shu-jie Li, Nir Barnea, Dong Bai, Hao-yu Shang, Niu Wan, Bao-jun Cai, and Chun-jian Zhang are gratefully acknowledged. This work is supported by the National Natural Science Foundation of China under Grants No. 12325507, No. 12547102, No. 12147101, No. 123B1011, No. 12422509, and No. 12375121, the National Key Research and Development Program of China under Grant No. 2022YFA1604900, No. 2024YFA1612500. The computations in this research were performed using the CFFF platform of Fudan University. 

\bmhead{Data availability}
All the data supporting the findings in this work are available within the manuscript and any additional data are available from the corresponding authors upon reasonable request.

\bmhead{Code availability}
Inquiries about the code in this work will be responded to by the corresponding authors.

\bmhead{Author contributions}
P. Li performed the analytical and numerical simulations and prepared the figures. K. J. Sun, B. Zhou and G. L. Ma supervised the project and contributed to the interpretation of the results. All authors participated in discussions and contributed to the preparation of the manuscript.

\bmhead{Competing interests}
The authors declare no competing interests.

\bibliography{myref.bib}

\newpage
\clearpage
\begin{onecolumn}

\setcounter{section}{0}
\begin{center}

    {\large \bf Supplemental materials for ``Universal imprinting of short-range correlations in relativistic heavy-ion collisions''} \\[12pt]
\end{center}

\section*{\centering{\small Analytical results of cluster structure with NN correlations}} 

Figure~\ref{Fig:cluster}a--c displays the analytical results on the dependences of $\varepsilon_{2}$, $\varepsilon_{3}$ and the RMS radius on the parameters $D_1$ and $D_2$ based on the 4$\alpha$-cluster structure of $^{16}$O. Fig.~\ref{Fig:cluster}d presents the effective one-body density $\rho^{(1)}_{cor}$ for four $\alpha$-clusters inside $^{16}$O. When $\gamma>0$, the attractive correlation effect in the center causes a slight increase in central density, while the opposite effect occurs when $\gamma<0$. For $\alpha=3$, the ratio indicates a correlation effect of approximately 1$\%$, while for $\alpha=1$, the correlation effect is approximately 5$\%$. We compare the HIC-SRC effect on $c_{E}\{2\}$ and $c_{E}\{3\}$ in both the spherical nuclei and the clustering structures in Fig.~\ref{Fig:cluster}e and f. Because the uncorrelated one-body density already includes cluster effects, the SRC effect has little modifying impact on the cluster structure.
    
        \begin{figure*}[htbp]
		\centering
		\includegraphics[scale=0.68]{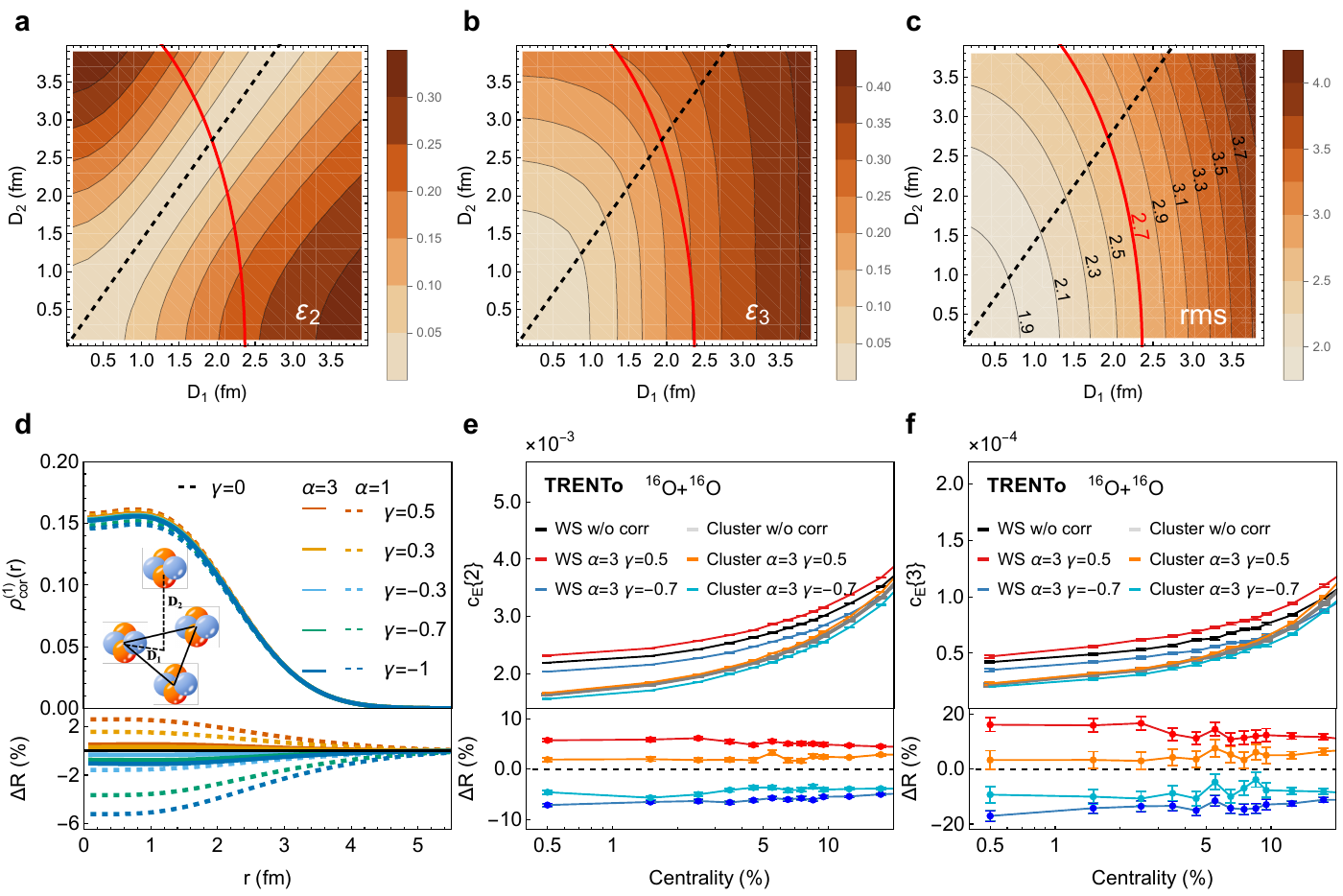}
		\caption{\textbf{Results of the shape and size of oxygen with NN SRCs.} \textbf{a}, \textbf{b} and \textbf{c}, The analytical results on the dependences of $\varepsilon_2$ (\textbf{a}), $\varepsilon_3$ (\textbf{b}) and RMS radius (\textbf{c}) on the parameters $D_1$ and $D_2$ for tetrahedral $^{16}$O, the solid red line represents a fit to the experimental RMS radius, while the black dashed line ($D_2/D_1=\sqrt{2}$) indicates a regular tetrahedral structure. \textbf{d}, Effective NN-correlated one-body density distributions $\rho^{(1)}_{cor}$ for four $\alpha$-clustered $^{16}$O with parameters $D_1=1.6$ fm and $D_2=3.5$ fm. \textbf{e} and \textbf{f}, The T\raisebox{-0.5ex}{R}ENTo results on $c_{E}\{2\}$ (\textbf{e}) and $c_{E}\{3\}$ (\textbf{f}), and its relative ratio $\Delta R$ for WS distribution and clustering distribution under leading order attractive correlations and repulsive NN correlations.}
	\label{Fig:cluster}
	\end{figure*}
    
    \begin{figure*}[htbp]
    \centering
	\includegraphics[scale=0.8]{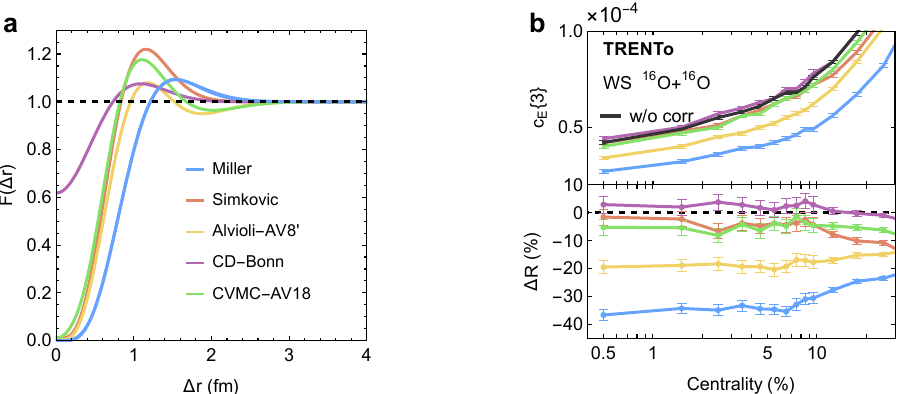}

    \caption{\textbf{Impact of different interactions in NN SRCs.} \textbf{a}, The two-point correlation function from different interactions. The results from Miller and Spencer~\cite{Miller:1975hu}, Simkovic~\cite{Simkovic:2009pp}, Alvioli--AV8$^\prime$~\cite{Alvioli:2009ab,Pudliner:1997ck}, CVMC--AV18~\cite{Cruz-Torres:2017sjy,Wiringa:1994wb}, CD--Bonn~\cite{Benhar:2014cka,Machleidt:1995km} models are shown for comparison. \textbf{b}, The T\raisebox{-0.5ex}{R}ENTo results on $c_{E}\{3\}$ and its relative ratio $\Delta R$ with different interactions.}
    \label{Fig:interaction}
    \end{figure*}

\section*{\centering{\small NN correlations with different interactions}}
Typical two-point correlation functions exhibit a short-range repulsive core at 0-1 fm and a weak attractive effect around 1 fm due to the tensor force, as shown in Fig.~\ref{Fig:interaction}a. The results from different interaction, including Miller and Spencer~\cite{Miller:1975hu}, Simkovic~\cite{Simkovic:2009pp}, Alvioli--AV8$^\prime$~\cite{Alvioli:2009ab,Pudliner:1997ck}, CVMC-AV18~\cite{Cruz-Torres:2017sjy,Wiringa:1994wb}, CD--Bonn~\cite{Benhar:2014cka,Machleidt:1995km}, are shown for comparison. We perform a parameter fitting on the two-point correlation functions corresponding to different interactions, except that the two-point correlation function for CVMC-AV18 is isospin-dependent, and the fitting parameters can be found in Ref~\cite{Cruz-Torres:2017sjy}. The CD--Bonn interaction has a relatively soft repulsive core, resulting in a weaker repulsive effect ($F(0)\approx0.6$) and the attractive part of the Simkovic interaction is stronger ($F(1\ \mathrm{fm})\approx1.22$) primarily driven by the tensor force. Consequently, in Fig.~\ref{Fig:interaction}b, the ratio of $c_{E}\{3\}$ in ultra-central collisions with CD--Bonn and Simkovic is slightly greater than the results of CVMC-AV18. In the Alvioli--AV8$'$ interaction, the attractive part is weaker, $F(1.2\ \mathrm{fm})\approx1.07$, while the Miller interaction has a longer repulsive range. The average correlation effect of these interactions is a very strong repulsive effect. Minute changes in the correlation function caused by different interactions can be reflected in the quantities with higher-order correlations. The relative difference in $c_{E}\{3\}$ exhibited by different interactions can reach 40$\%$. This suggests that the NN correlation effect in heavy-ion collisions is potentially a powerful probe for testing nuclear interactions.

    \begin{figure}[htbp]
    \centering
	\includegraphics[scale=0.8]{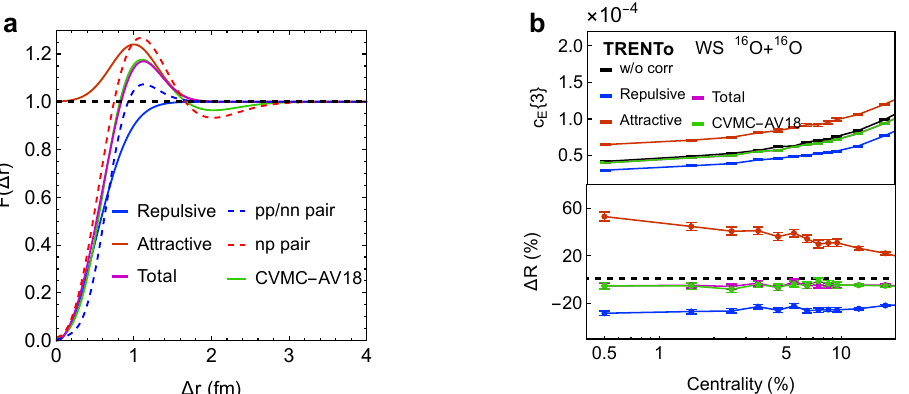}

    \caption{\textbf{Impact of isospin-dependent interaction in NN SRCs.} \textbf{a}, The two-point correlation function for CVMC-AV18 interaction (solid green line) and its repulsive part (solid blue line) and attractive part (solid red line). The dashed red line and the dashed blue line represent the two-point correlation functions for the $np$ pairs and the $pp$/$nn$ pairs, respectively. \textbf{b}, The T\raisebox{-0.5ex}{R}ENTo results on $c_{E}\{3\}$ and its relative ratio $\Delta R$ with isospin-dependent CVMC-AV18 interaction(solid green line), repulsive part (solid blue line), attractive part (solid red line) and the sum of two parts (solid purple line).}
    \label{Fig:poandne}
    \end{figure}
    
\section*{\centering{\small NN correlations with isospin dependence}}
Extensive experimental evidence from a wide range of nuclei, from light to heavy (e.g. $^{12}$C, $^{56}$Fe and $^{208}$Pb), shows that the number of neutron-proton ($np$) SRC pairs far exceeds that of proton-proton ($pp$) or neutron-neutron ($nn$) pairs (about 20 times in $^{12}$C)~\cite{Subedi:2008zz}. It strongly supports the dominant role of the tensor force component~\cite{Sargsian:2005ru,Schiavilla:2006xx}, which is most significant in the $np$ pair channel while being suppressed in the $pp$/$nn$ pairs due to the inherent Pauli exclusion principle.

Theoretically, the two-point correlation function is predicted to exhibit a strong isospin dependence that the attractive part of the correlation function in $np$ pairs is stronger than that in $pp$/$nn$ pairs, while the repulsive part is weaker due to the lack of the Pauli exclusion principle. We employ an adaptive Monte Carlo method to generate isospin-dependent correlated samples. In Fig.~\ref{Fig:poandne}a, $pp$/ $nn$ pairs and $np$ pairs are sampled using distinct correlation functions, represented by the red and blue dashed lines, respectively. We also implement a phenomenological approach that decomposes the correlation function into separate attractive (solid red line) and repulsive (solid blue line) components. The correlation terms from both parts can be directly added to reproduce the realistic two-point correlation function. This parameterization method has no isospin dependence and represents an average correlation effect.

Based on the results in Fig.~\ref{Fig:poandne}b, although the true two-point correlation function exhibits isospin dependence, the initial-state quantities of heavy-ion collisions fail to demonstrate isospin-dependent correlation properties when compared to the average two-point correlation function. It reflects that the current quantities are all influenced by the average isospin correlation effect. Consequently, as long as the correlation functions exhibit the same average effect, the final impact on the observables will also be identical. This arises from our initial-state modeling assumption of an isospin-independent nucleon-nucleon cross section. It also points to a possible new direction for improving phenomenological models for heavy-ion collisions.
   
It is noteworthy that the attractive and repulsive effects in the realistic two-point correlation function mutually cancel. In contrast, this is not the case for the correlation functions in the currently popular NLEFT and PGCM models~\cite{Zhang:2024vkh}, as they have only a single central attractive or repulsive effect. Neither model well captures the abrupt change in the correlation function observed at $\sim$ 1.0 fm. 

\end{onecolumn}

\end{document}